\documentclass[prd]{revtex4}
\usepackage{amsmath}
\usepackage{array}
\usepackage{dcolumn}
\usepackage{longtable}
\usepackage[latin1]{inputenc}
\usepackage[dvips]{graphicx}

\newcommand{\ep}{\epsilon}
\newcommand{\be}{\begin{equation}}
\newcommand{\ee}{\end{equation}}
\newcommand{\ber}{\begin{eqnarray}}
\newcommand{\eer}{\end{eqnarray}}

\newcommand{\p}{\partial}

\def\case#1/#2{\textstyle\frac{#1}{#2} }
\newcommand{\bra}[1]{\left(#1\right)}
\newcommand{\bras}[1]{\left[#1\right]}
\newcommand{\brac}[1]{\left\{#1\right\}}
\newcommand{\reff}[1]{(\ref{#1})}

\begin{document}

\title{Harmonic generation of gravitational wave induced Alfvén waves}
\author{Mats Forsberg and Gert Brodin}
\affiliation{Department of Physics, Ume\aa\ University, SE-901 87 Ume{\aa},
Sweden} 
\date{\today}

\begin{abstract}
Here we consider the nonlinear evolution of Alfvén waves that have
been excited by gravitational waves from merging binary pulsars. We derive a
wave equation for strongly nonlinear and dispersive Alfvén waves. Due to the
weak dispersion of the Alfvén waves, significant wave steepening can occur,
which in turn implies strong harmonic generation. We find that the harmonic
generation is saturated due to dispersive effects, and use this to estimate
the resulting spectrum. Finally we discuss the possibility of observing the
above process.

\end{abstract}

\maketitle

\section{Introduction}

The launch of large projects for the detection of gravitational waves such as
LIGO (Laser Interferometer Gravitational Wave Observatory) together with
ambitious projects under development such as LISA (Laser Interferometer
Space Antenna) \cite{maggiore} has increased the hope for successful
detection of gravitational waves (GW:s) during the next few decades, and
stimulated much work (e.g. Refs. \cite{maggiore,Schutz1999}). Naturally the
research devoted to detection concerns low amplitude GW:s. In an
astrophysical context close to the GW source, the gravitational waves can
propagate in a plasma medium, and the amplitudes are larger. In Refs. \cite
{brodinmarklund,Papadouplous2001,bms,bmd2,Balakin2003,Servin2000,ignatev,kallberg2004} the authors have studied
nonlinear responses to the gravitational wave by the plasma medium, although
the backreaction has been neglected. The nonlinear response gives raise to
effects such as parametric instabilities \cite
{Papadouplous2001,bmd2,Balakin2003,Servin2000}, large density fluctuations 
\cite{bms,ignatev}, photon acceleration \cite{bms} and wave collapse \cite
{kallberg2004}. The application of gravitational wave processes to
astrophysics has been discussed by for example Refs. \cite
{bmd1,Moortgat2003,Mosquera2002,Isliker,Moortgat2006}, and to cosmology by Refs. \cite
{Papadoupolus2002,MDB2000,Hogan2002,kuiroukidis}. A number of works studying nonlinear
propagation of gravitational waves including the backreaction from the
plasma have also been written, see e.g. Refs. \cite
{ignatev,Balakin2003,servinbrodin}.

In Ref. \cite{kallberg2004} the coupled evolution of GW:s and
(compressional) Alfvén waves were considered, and a condition for nonlinear wave
collapse was derived. Here we develop that work, taking into account the
dispersive properties of the Alfvén waves. For conditions, when wave collapse do
\textit{not} occur, we here show that the nonlinear evolution
lead to significant wave steepening, and associated high harmonic generation.
Furthermore, we find that the harmonic generation is saturated due to the
dispersive effects associated with the Hall current. As a result, we are
able to estimate the resulting wave spectrum. Due to the strong harmonic
generation, it turns out that electromagnetic wave frequencies several
orders of magnitudes larger than the original GW frequency can be generated.
Finally we discuss the possibility that such radiation might be observable with satellite
based radio arrays \cite{ALFA,SIRA}.

\section{Coupled Alfvén and Gravitational waves}

The metric of a linearized gravitational wave propagating in the $z$%
-direction can be written as \cite{Landau-Lifshitz} 
\begin{equation}
\mathrm{d}s^{2}=-\mathrm{d}t^{2}+[1+h(z - ct)]\mathrm{d}x^{2}+[1-h(z - ct)]\mathrm{d}%
y^{2}+\mathrm{d}z^{2}\ \label{lineelement},
\end{equation}
where we have assumed linear polarization, as will be justified below. For an
observer comoving with the time coordinate, the natural frame for
measurements is given by 
\begin{equation}
	e_{0}=\partial _{t}\ , \
	e_{1}=\left( 1-{\tfrac{1}{2}}h\right) \partial _{x} \ , \
	e_{2}=\left( 1+{\tfrac{1}{2}}h\right) \partial _{y} \ , \
	e_{3} = \partial_{z}\ . \label{tetrad}
\end{equation}
We now introduce the 3-vector notation such that ${\boldsymbol{\nabla}}\equiv (e_{1},e_{2},e_{3})$,
$\boldsymbol{E} \equiv (E^1,E^2,E^3)$ is the electric field and $\boldsymbol{B} \equiv (B^1,B^2,B^3)$ is the magnetic field. It can be shown \cite{bmd1} that in the frame \reff{tetrad} Maxwell's equations can be written 
\begin{eqnarray}
		{\boldsymbol{\nabla\cdot}}{\boldsymbol{E}} &=&\frac{\rho}{\varepsilon_{0}}\ , \label{eq:max1} \\
		{\boldsymbol{\nabla\cdot}}\mathbf{B} &=&0 \ ,  \label{eq:max2} \\
		\p_t {\boldsymbol{E}}-c^2{\boldsymbol{\nabla\times}}{\boldsymbol{B}}
		&=& -\frac{{\boldsymbol{j}}_{\!_{E}}}{\varepsilon_0}-\frac{{\boldsymbol{j}}}{\varepsilon_0}\ , \label{eq:max3} \\
		\p_t{\boldsymbol{B}} + {\boldsymbol{\nabla\times}}{\boldsymbol{E}} &=&-\frac{{\boldsymbol{j}}_{\!_{B}}}{c \varepsilon_0} \ . \label{eq:max4}
\end{eqnarray}
Here
$\rho\equiv \sum_s q \gamma n$ is the charge density, $q$ and $n$ denotes the charge and the proper particle density for a particle of species $s$ and ${\boldsymbol{j}}_{\!_{E}}$ and ${\boldsymbol{j}}_{\!_{B}}$ are the effective gravitational current densities, defined by
\begin{eqnarray}
	j_{\!_{E}}^{1} &=&j_{\!_{B}}^{2} \equiv {\frac{c \varepsilon_0}{2 }}\bra{ E^{1} - c B^{2} } \p_z h \ ,  \label{EffectiveA} \\
	j_{\!_{E}}^{2} &=&-j_{\!_{B}}^{1} \equiv -{\frac{c \varepsilon_0}{2}}\bra{ E^{2} + c B^{1} } \p_z h \ .  \label{EffectiveB}
\end{eqnarray}
Next we assume the presence of a background magnetic field, $\mathbf{B}_{0}=B_{0}\mathbf{e}_{1}$, and introduce perturbations such that $n=n_{0}+\delta n$, $\mathbf{B} = (B_{0}+B_{x})\mathbf{e}_{1}$, $\mathbf{E}=E_{y}\mathbf{e}_{2}$ and $\mathbf{v}=v_{y}\mathbf{e}_{2}+v_{z}\mathbf{e}_{3}$, corresponding to compressional Alfvén waves excited by the GW:s.
We also note from Ref. \cite{bms} that in the case of gravitational waves propagating in a magnetized plasma, with the magnetic field perpendicular to the direction of propagation, only the linear component of the GW polarization considered in Eq. \reff{lineelement} couples effectively to the electromagnetic wave.
Furthermore $v_{y}\ll c$, (as $ v_{y}/c \sim h $), and we therefore neglect terms of the type $v_{y}^{2}$ and $v_{y}h$, but allow for $v_{z}\sim c$.
We will also consider slow variations such that $\partial _{t}\ll \omega _{c}\equiv qB_{0}/m$
for each plasma species.

Following Ref. \cite{kallberg2004}, letting the variables depend on $z$ and $t$, we find that the Maxwell and fluid equations can be
reduced to 
\begin{eqnarray}
	\partial _{t}E_{y}- c^2 \partial _{z}B_{x} + \sum_{s} \frac{q}{\varepsilon_0}\gamma (n_{0}+\delta n)v_{y} 
	&=&-{\frac{1}{2}}E_{y}\partial _{t}h + {\frac{c^2}{2}}(B_{0}+B_{x})\partial_{z}h \ ,  \label{ampeq} \\
	\partial _{t}B_{x}-\partial _{z}E_{y} &=&
	-{\frac{1}{2}} E_{y}\partial _{z}h +{\frac{1}{2}}(B_{0}+B_{x})\partial_{t}h \ ,  \label{faraday} \\
	\partial _{t}\left( \gamma (n_{0}+\delta n)\right) &=&-\partial _{z}\left(
	\gamma (n_{0}+\delta n)v_{z}\right)  \label{cont} \ , \\
	\partial _{t}(\gamma v_{y})+v_{z}\partial _{z}(\gamma v_{y}) &=&{	\frac{q}{m}}(E_{y}+v_{z}(B_{0}+B_{x})) \ ,  \label{fleq2} \\
	\partial _{t}(\gamma v_{z})+v_{z}\partial _{z}(\gamma v_{z}) &=&-{ \frac{q}{m}}v_{y}(B_{0}+B_{x}) \ ,  \label{fleq3} 
\end{eqnarray}
where Eqs. \reff{cont}-\reff{fleq3} holds for each particle species. However, for notational 
convenience we omit the index denoting species, as we at this stage want 
to cover both the case of an electron-ion plasma as well as an electron-positron plasma.
The above system of Eqs. \reff{ampeq}-\reff{fleq3} should be complemented by the non-trivial part of the linearized Einstein field
equations \cite{bms, servinbrodin, grishchuk}
\be
	\bra{ c^{-2} \p_{t}^{2} - \p_{z}^{2} } h =\kappa \bra{\delta T_{11}-\delta T_{22}} \ ,  \label{ee1122} 
\ee
where $\kappa \equiv 8 \pi G / c^4$, $G$ is Newton's gravitational constant and $\delta T_{ij}$ are the perturbed components of the energy momentum tensor. Provided the coupled Alfvén waves and GW:s move together close to the speed of light, the vacuum expressions for the effective currents \reff{EffectiveA}-\reff{EffectiveB} holds approximately. The above system of equations has previously been studied by Ref. \cite{bms} and Ref. \cite{kallberg2004}. In particular Ref. \cite{kallberg2004} derived a system of two coupled equations of the form 
\begin{eqnarray}
	(\p_{t}+\mathcal{U}(B_{x})\p_{z})B_{x} &=&\frac{1}{2}B_{0}\partial _{t}h \ ,  \label{waveq} \\
	\bra{\frac{1}{c^2}\p_{t}^{2} - \p_{z}^{2}}h &=&-2 \kappa \frac{ B_{0}B_{x}}{\mu_0} \ , \label{eefinal}
\end{eqnarray}
where $\mathcal{U}(B_{x})=c-(c^3/2C_{A}^{2})\left( B_{0}/\left(
B_{0}+2B_{x}\right) \right) ^{3/2}$, 
$C_{A} \equiv 
c(\sum_{s}\omega _{p}^{2}/\omega _{c}^{2})^{-1/2}$ is the non-relativistic Alfvén velocity 
and $\omega_p = (n_0 q^2/\ep_0 m)^{1/2}$ is the plasma frequency for each species. 
%
%
In contrast to Refs. \cite{bms,kallberg2004} we will here concentrate on the evolution of the Alfvén
waves at a distance further from the gravitational source, where the Alfvén-GW coupling is of less importance, and the GW source terms for the Alfvén waves can be neglected. Thus from now on we will assume the perturbations to be of the
form $\psi \approx \psi (z-V_{A}t)$, where $V_{A} = C_A / \sqrt{1+C_A^2/c^2}$ is the (roughly constant) relativistic Alfvén velocity, which allows us to use $\partial _{t}\approx - V_{A}\partial _{z}$. With these approximations we can reduce the system (\ref{ampeq})-(\ref{fleq3}) to 
\begin{eqnarray}
		(\p_{t}^{2}-c^{2}\p_{z}^{2})B_{x}+\sum_{s}\frac{mn_{0}}{\varepsilon _{0}}V_{A}\p_{z}
		\left[ \frac{\p_{z}(\gamma v_{z})}{B_{0}+B_{x}}\right] &=&0 \ ,  \label{wave3} \\
		(v_{z}-V_{A})\p_{z}(\gamma v_{y})-\frac{q}{m}[v_{z}(B_{0}+B_{x})-V_{A}B_{x}] &=&0 \ ,  \label{fleq2_2} \\
		(v_{z}-V_A)\p_{z}(\gamma v_{z}) + {\frac{q}{m}}v_{y}(B_{0}+B_{x}) &=& 0 \ . \label{fleq3_2}
\end{eqnarray}
Since $\omega_c^{-1} \p_t \ll 1$ and $\gamma v_y/c \ll 1$, Eq. \reff{fleq2_2} can be used to obtain an approximate expression of  $v_z$,
\begin{equation}
	v_{z}=V_{A}\frac{B_{x}}{B_{0}+B_{x}} \ ,  \label{v_z0}
\end{equation}
which in turn can be used together with \reff{wave3} to express the wave equation as
\begin{equation}
	(\p_{t}^{2}-c^{2}\p_{z}^{2})B_{x} + \sum_{s}\frac{mn_{0}}{\varepsilon _{0}}V_{A}^{2}\p_{z}
	\frac{B_{0}\p_{z}B_{x}}{\left[B_{0}^{2}+2B_{0}B_{x}+(1-V_{A}^{2}/c^{2})B_{x}^{2}\right] ^{3/2}}=0 \ . \label{wave4}
\end{equation}
By using $\p_{t}^{2}-c^{2}\p_{z}^{2}\approx -(c+V_A)(\p_t + c \p_z)\p_z$ we obtain 
\begin{equation}
	\p_{t} B_{x} + \mathcal{V}(B_{x})\p_{z}B_{x}=0 \ , \label{wave_v_z0}
\end{equation}
where 
\begin{equation}
		\mathcal{V}(B_{x})=\brac{ c - \bra{c-V_A}\bras{1+2\frac{B_x}{B_0} + \bra{1-\frac{V_A^2}{c^2}}\frac{B_x^2}{B_0^2}}^{-3/2}} \ .
		\label{wavespeed_v_z0}
\end{equation}
We note here that the velocity $\mathcal V(B_x)$ agrees with the velocity $\mathcal U(B_x)$ in Eq. \reff{waveq} (used previously by Ref. \cite{kallberg2004}) to first order in $c/C_A$. Expression \reff{wave_v_z0} is slightly more accurate, however, as $c/C_A \ll 1$ has not been used in the later derivation.

Making the transformation $t=\tau \ , \  z=\zeta + \int_0^\tau \mathcal V (B(\zeta,\tau')) d\tau'$ we have $\p_t \rightarrow \p_\tau - (\mathcal V(B_x)/R) \p_\zeta$ and $\p_z \rightarrow (1/R) \p_\zeta$, where $R \equiv 1 + \int_0^\tau (\p \mathcal V(B_x)/\p \zeta) d\tau'$. As a consequence the general solution of Eq. \reff{wave_v_z0} can be written $B_x(z,t)=B_x(\zeta)$, which formally shows that wave steepening continues until wave breaking occur, such that $R \rightarrow 0$ and $\p_z \rightarrow \infty$ at some point for a finite time. Thus, in order to include dispersive effects that will prevent wave breaking we will need a more accurate expression than Eq. \reff{v_z0} for $v_z$. Dispersive effects are then included by keeping terms of a higher order in an $\omega_c^{-1}\partial_t$-expansion. How to implement this is to some extent dependent on whether we study an electron-ion plasma (in which case it is the higher order expression for the ion polarization drift that first leads to wave dispersion), or whether we consider an electron-positron plasma (in which case electron and positron motion contributes to wave dispersion simultaneously). For the astrophysical applications to be considered below, an electron-ion plasma is more appropriate, and thus we will focus on this case in the remainder of the article, using index $i$ to denote ion quantities below. The only consequence for the calculations that has been made above is that we can simplify the expression for the Alfvén velocity slightly, such that
$C_A  = c(\sum_{s}\omega _{p}^{2}/\omega _{c}^{2})^{-1/2} = (B_0^2/\mu_0 m_i n_0)^{1/2}$. Next, using Eq.  \reff{v_z0} as an  approximation of $v_{zi}$ in Eq. \reff{fleq3_2} gives
\begin{equation}
		v_{yi}=\frac{m_i}{q_i}\left( \frac{B_{0}^{2}V_{A}^{2}}{B_{0}+B_{x}}\right)
		\frac{\p_{z}B_{x}}{\left[ B_{0}^{2}+2B_{0}B_{x}+\left( 1-\frac{V_{A}^{2}}{c^{2}}\right) B_{x}^{2}\right] ^{3/2}} \label{v_y} \ ,
\end{equation}
which can be inserted into Eq. \reff{fleq2_2}, resulting in a more accurate expression for $v_{zi}$,
\begin{equation}
		v_{zi} = \frac{V_A B_x}{B_0+B_x} - \frac{V_{A}^{3}B_{0}^{3}}{\bra{ B_{0}+B_{x}}^{2}}\bra{\frac{m_i}{q_i}}^{2}
		\left\{ \frac{\p_{z}^{2}B_{x}}{\bras{B_{0}^{2}+2B_{0}B_{x}	+	\bra{ 1-\frac{V_{A}^{2}}{c^{2}} } B_{x}^{2}}^{2}}
		- \frac{4\bras{ B_{0}+\left( 1-\frac{V_{A}^{2}}{c^{2}}\right)B_{x}}	\brac{ \p_{z}B_{x}}^{2}}{\bras{ B_{0}^{2}+2B_{0}B_{x}
		+ \left( 1-\frac{V_{A}^{2}}{c^{2}}\right) B_{x}^{2} }^{3}}\right\} \ .  \label{v_z}
\end{equation}
This corrected expression of $v_{zi}$ combined with Eq. \reff{wave3} results in a wave equation of the form
\be
		\p_t B_x + \mathcal V (B_x) \p_z B_x + \frac{c-V_A}{ B_0 + B_x }\bra{\frac{V_A}{\omega_{ci}}}^{2}
		\p_z \brac {\frac{B_{0}^{7}}{\bra{ B_{0}+B_{x}}}
		\bras{ \frac{\p_{z}^{2}B_{x}}{g^5(B_x)}	- \frac{4\left[ B_{0}+\left( 1-\frac{V_{A}^{2}}{c^{2}}\right)B_{x}\right]
		\left( \p_{z}B_{x}\right)^{2}}{g^7(B_x)} } } = 0 \ , \label{full_evolution}
\ee
where $\mathcal{V}(B_x)$ is given by \reff{wavespeed_v_z0} and the auxiliary function $g(B_x)$ is defined by
\be
		g(B_x) \equiv \left[B_{0}^{2}+2B_{0}B_{x}	+	\left( 1-\frac{V_{A}^{2}}{c^{2}}\right) B_{x}^{2}\right]^{1/2} \ . \label{g(B_x)}
\ee
As we can see, Eq. \reff{full_evolution} now describes the fully nonlinear evolution combined with dispersive effects.
In the weakly nonlinear limit this reduces to the celebrated Korteweg de Vries (KdV) equation,  
\be
		\p_{t}B_{x}+V_{A}\p_{z}B_{x} + 3 (c - V_A) \frac{B_{x}}{B_{0}}\p_{z}B_{x}
		+ (c - V_A) \frac{V_{A}^{2}}{\omega_{ci}^{2}}\p_{z}^{3}B_{x}=0  \ .  \label{KdVDR}
\ee
As is wellknown \cite{doddeilbeck}, for vanishing boundary conditions the general solution to the KdV equation involves a train of solitons, where the steepening effects due to nonlinearity is balanced by the dispersive effects due to the last term of Eq. \reff{KdVDR}.
Our main interest here is the steepening of an initially sinusoidal wave
profile. A study of the KdV equation for such initial conditions shows that
the wave steepening induced by the nonlinearity leads to a harmonic content. However, as we are interested in generation of \textit{large} harmonic content, with wave steepening occurring on a comparatively fast time scale, we need to study the fully nonlinear dispersive Eq. \reff{full_evolution} rather than Eq. \reff{KdVDR}.

%
%
An initially sinusoidal wave described by Eq. \reff{full_evolution} can at first be approximated with Eq. \reff{wave_v_z0} if the initial wave frequency $\omega$ fulfills $\omega \ll \omega_{ci}$. As a consequence the sinusoidal profile will undergo wave steepening until dispersive effects becomes important leading to a saturation of the steepening and the approach to a steady state profile, as can be described by Eq. \reff{full_evolution}. Thus we are looking for solutions to Eq. \reff{full_evolution} that are periodic and static in a frame moving with the wave. The general methods for finding solutions of this type is described in some detail in Ref. \cite{Shukla}. First we let $\partial_t \rightarrow -V_{An}\partial_z$, where $V_{An}$ is 
a constant representing the nonlinearly modified Alfvén velocity. Integrating the resulting equation with respect to $z$, we get an 
equation that are analogous to a particle moving in a potential \cite{Shukla}. For appropriate choices of integration constants, the motion corresponds to a particle moving in a potential well, in which case we obtain solutions of the desired type, i.e. $B_x=B_x(z-V_{An}t)$, where $B_x$ is a periodic function of the argument. The resulting differential equation for $B_x$ has to be solved numerically. The degree of wave steepening in the resulting wave profile depends on the amplitude of the oscillations as well as on $\omega/\omega_{ci}$, parameters which are related to the integration constants. For given integration constants, the corresponding physical parameters can be calculated straightforwardly, and thereby solutions with parameters of physical interest can be found by some trial and error. Solutions of astrophysical relevance showing significant wave steepening corresponding to large harmonic content of the wave profile will be presented in the preceding section.
%
%
\section{Astrophysical example}

In this section we are going to investigate the nonlinearly modified 
Alfvén waves, and to what extent these can be observed. Based on the fact that 
electromagnetic signals are easier to detect than gravitational
ones of the same power, we will be looking for a spectral signature of 
GW-generated Alfvén waves. As is wellknown, the gravitational sources 
have high powers only up to frequencies of the order $\sim 
1\mathrm{kHz}$, which is much below the
lowest observable radio frequencies \cite{ALFA,SIRA}. Thus for GW 
induced radio wave detection to be
possible, we will need a mechanism to convert some fraction of the
electromagnetic energy to higher frequencies. As we will demonstrate below,
the processes studied in Section II can provide the basis for such
mechanisms, and thus lead to a large increase of the frequency for a
significant fraction of the electromagnetic spectrum.

\begin{figure}[htbp]
		\includegraphics{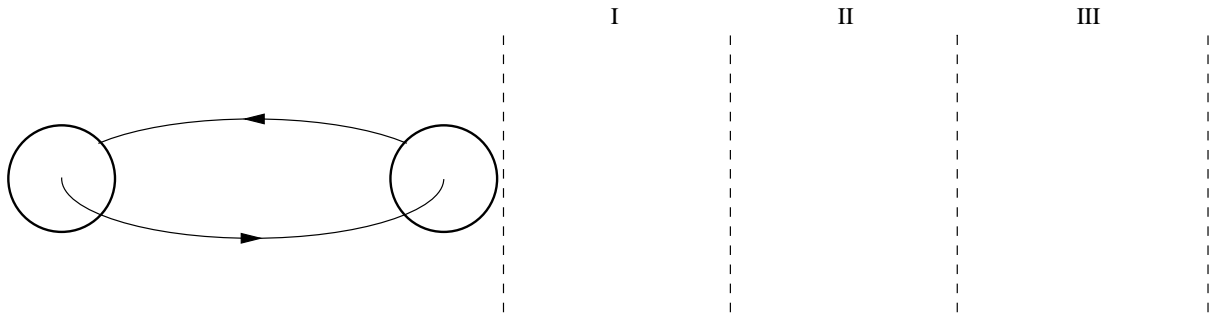}
\caption{The neighborhood of the binary system is divided into three regions: region I ($10 R_S-30 R_S$), the energy conversion zone; region II ($30 R_S - 3500 R_S $), the nonlinearity enhancement zone and region III ($3500 R_S - 7000 R_S$), the wave steepening zone.} \label{binary}
\end{figure}

Below we will study a concrete example. As a source of gravitational
radiation we consider a binary system. At least one of the objects should
have a strong magnetic field (in order to make the Alfvén phase velocity close
to $c$), and the objects should be compact (as to make the gravitational
wave frequency and amplitude before merging reasonably large). For definiteness we
study a system consisting of two neutron stars of equal mass $M_{\odot }$,
separated by a distance of $20\ R_{S}$, where $R_{S}=2GM_{\odot
}/c^{2}\approx 3$ km. Furthermore, the surface magnetic field of each
neutron star is assumed to be $4 \times 10^{6}$ T. The surroundings of the binary
system can loosely be divided into three regions (Fig.\ \ref{binary}),
depending on which physical mechanisms that is dominating.

\subsection{Energy conversion zone}

The interval $10\ R_{S}-30\ R_{S}$ from the center of mass (CM) roughly
constitutes region I, which is the region where most of the gravitational
energy is gained by the Alfvén-wave. Using a Newtonian approximation, with $%
d=\alpha R_{S}$ and $r=\beta R_{S}$, it is straightforward to show that $%
|h|\sim (2\alpha \beta )^{-1}$, where $d$ is the separation distance between
the binary objects and $r$ is the observation distance from the center of
mass of the system. In order to obtain an estimate of the amplitude of the
generated EM-wave, we note that in the near zone (i.e. where the magnetic
pulsar magnetic field decays as a dipole) the plasma density is low, and we
expect that for our purposes here, we can approximate the medium as vacuum.
A calculation of the electromagnetic amplitude generated by the GW during
such circumstances has been done by Ref. \cite{bms}. Combining the
above expression for the gravitational wave amplitude at given distances
with Eq.\ (14) of Ref. \cite{bms}, the GW-induced magnetic field
amplitude $\delta B$ can thus be estimated. The result is 
\begin{equation}
		\frac{\delta B}{B_{0}}\sim 1.8 \times 10^{-4} \ ,
\end{equation} 
at the of end region I.

\subsection{Nonlinearity enhancement zone}

In region II (approximate interval $30\ R_{S}-3500\ R_{S}$ from the CM)
the ratio determining the degree of nonlinearity, $\delta B/B_0$, is increasing quadratically
. The reason is that $\delta B$ suffers spherical attenuation (due to the high frequency) whereas $B_0$ decays as a dipole field until the light cylinder is reached. The outer bound of region II, with a light cylinder at $3500 R_S$, is somewhat artificially chosen, and correspond to a pulsar period of $35$ ms. Note, however, that the nonlinearity enhancement mechanism is sufficient to reach the strongly nonlinear regime (i.e. $\delta B/B_0 \sim 1 $) within region II even for faster pulsars.

\subsection{Wave steepening zone}

In region III (approximate interval $3500\ R_{S}-7000\ R_{S}$), the amplitude of the Alfvén waves is strongly nonlinear, and thus pronounced wave
steepening will occur here, as described by Eq. \reff{KdVDR}. The characteristic propagation length $L_{p}$ for large steepening to occur is of the order 
\begin{equation}
	L_{p}\sim \frac{\lambda }{2}\frac{V_{A}}{\delta V_{A}} \ ,  \label{steepening_length}
\end{equation}
where $\delta V_{A}$ is the velocity difference between the maximum and minimum wave velocity within the amplitude wave profile with dispersive effects neglected. At the beginning of region III we have $B_{0}\simeq 3 \times 10^{-3} \mathrm{T}$. Assuming a moderate plasma density, of the order $n_{0}\simeq 10^{10}\mathrm{m}^{-3}$, gives an Alfvén velocity $V_{A}\simeq 0.9c$. Furthermore, a relatively strong nonlinearity with $\delta B / B_0 = 0.25$, consistent with the estimate made in section III B, correspond to $\delta V_{A}\simeq 0.2c$, which combined with a wavelength $\lambda \simeq 1.7 \times 10^{6}\mathrm{m}$ result in a steepening distance located reasonably well within region III.

\begin{figure}[htb]
		\includegraphics[width={12cm}]{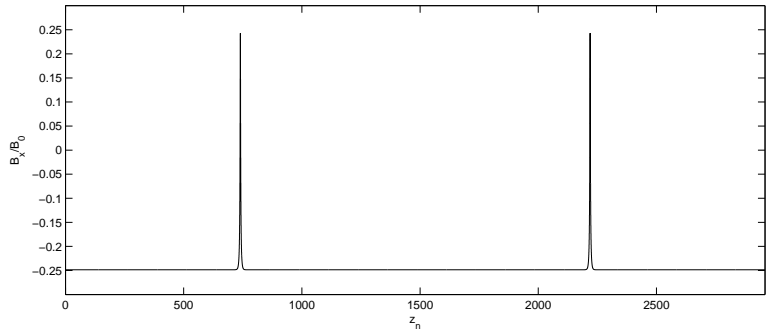}
\caption{The wave profile for a moderately nonlinear amplitude ($\delta 
B/B_0 \approx 0.25 $) as a function of the normalized distance $z_n = z \ \omega_{ci}/V_{A}$. The wavelength $\lambda_n \approx 1500$ correspond to a ratio $\omega / \omega_{ci}\approx 1/240 $.} \label{magfield}
\end{figure}

\begin{figure}[htb]
		\includegraphics[width={12cm}]{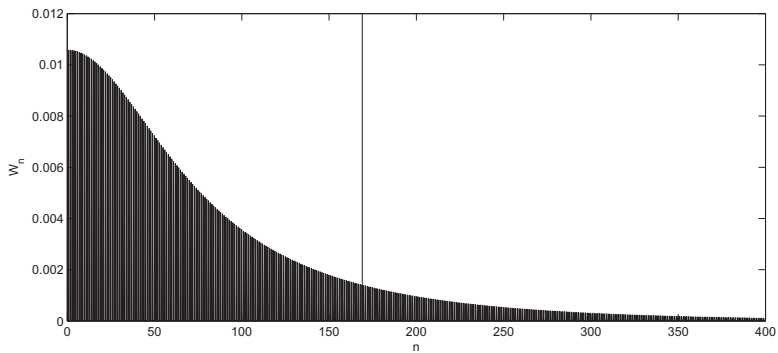}
\caption{The normalized energy density (total energy density = 1) content $W_n$ of the wave profile as a function of the 
harmonic number $n$ corresponding to the wave profile in Fig 2. The vertical line around $n\approx 170$  separates the harmonics exceeding the lower sensitivity  bound of SIRA from the frequencies that are too low to be detected. For our given example, the amount of energy that can be detected is of the order of $ 10 \%$ of the total wave energy.} \label{spectra}
\end{figure}

Due to the nonlinear velocity dependence of the amplitude, wave steepening continues until the dispersive effects become important. As 
the initial GW-generated angular frequency is $\omega \approx 1.1\times 10^3 \mathrm{s}^{-1}$, whereas $\omega_{ci} \approx 2.7 \times 10^5 \mathrm{s}^{-1}$ in our case, and the amplitude is strongly nonlinear, the steady state profiles are dramatically changed as compared to the initially sinusoidal wave profiles. The steady state profile derived numerically from Eq. \reff{KdVDR} corresponding to $\omega / \omega_{ci} = 1/240$ and $\delta B / B_0 \approx 0.25$ is shown in Fig. \ref{magfield}. Since, the temporal and spatial derivatives are orders of magnitudes larger than the corresponding values for the initial profile, the spectral content has changed dramatically from the initially quasi-monochromatic wave. The spectrum corresponding to the same data as in Fig. \ref{magfield} is shown in Fig. \ref{spectra}. There it is seen that the highest harmonic generation of Alfvén waves may approach the 100 kHz range. Unfortunately, this is still not enough to reach the radio window for earth based antennas, where the lower cut-off lies around $20\mathrm{MHz}$. However, proposals such as the Astronomical Low Frequency Array (ALFA) \cite {ALFA} and more recently the Solar Imaging Radio Array (SIRA) \cite{SIRA} show that satellite based antennas can provide an opening for observation, since such arrays are planned for a sensitivity down to frequencies of the order $30\mathrm{kHz}$. As shown in Fig. \ref{spectra}, the spectral content exceeding this frequency is of the order $10 \% $ of the total wave energy density. Thus the astrophysical processes outlined here in section III opens for the possibility to correlate and interpret GW-observations made by LIGO or LISA with radio wave observations made with SIRA.

\section{Summary and Discussion}
We have considered the coupling of GW:s and Alfvén waves in a magnetized plasma. The
dependence of the Alfvén velocity on the magnetic field amplitude makes
the evolution equation for the Alfvén waves strongly nonlinear. As described
in section II, GW-generated Alfvén waves are subject to nonlinear wave steepening,
which is directly associated with harmonic generation. In order to find a
saturation mechanism for the steepening process, we have included dispersive
effects in our model by solving the momentum equation to third order in an $%
\p_t /\omega _{c}$-expansion. As demonstrated in section IIIc, the high
amplitude steady state solutions for the Alfvén waves show strong harmonic
generation, which for sufficient amplitude does not saturate until the
frequencies of the harmonics approach the cyclotron frequency of the ions.

As a specific example we have considered a case where the GW:s associated with a
neutron star-pulsar merging generate strongly nonlinear Alfvén waves,
resulting in high harmonic generation of the Alfvén waves. As shown, the
original GW-frequencies of the order $\sim 200\mathrm{Hz}$\ may
generate signals with harmonic number $n \sim 200$ or even larger. 
Unfortunately this is not sufficient to reach the radio window for
earth based detection, but proposals for satellite based radio observations
\cite{SIRA}, that can detect radio signals down to $\sim 30\mathrm{%
kHz}$ could make observations possible. This opens for the future
possibility to correlate GW-observations made by for example LISA \cite{maggiore} with
radio wave observations made with SIRA \cite{SIRA}, to make detailed comparison
with theories for the coupled electromagnetic-gravitational evolution. The process described here can be even more effective than in our example, if pulsars with stronger magnetic fields are considered.

Finally we should point out that for the Alfvén waves generated in our
example to travel through interstellar distances, they must be more or less
decoupled from the medium to become ordinary radio waves. For this to
happen, the density should fall off with distance sufficiently fast after
the wave steepening region in the previous example, such that the plasma
frequency falls beyond the highest parts of the wave frequency before the
cyclotron frequency does so. In our example this is fulfilled, since
the highest harmonic frequencies can exceed the plasma frequency
already within the wave steepening region. However, for an accretion disc
that is still thick at a distance when the magnetic field falls below the
wave frequency, cyclotron damping could limit the possibilities for Radio
wave detection to a significant extent.


\section{References}

\begin{thebibliography}{99}

\bibitem{maggiore}  M. Maggiore, Phys. Rep. \textbf{331}, 283 (2000); See
also URL http://www.ligo.caltech.edu/ and URL http://lisa.jpl.nasa.gov/



\bibitem{Schutz1999}  B. F. Schutz, Class. Quantum Grav. \textbf{16}, A131
(1999)


\bibitem{bms}  G. Brodin, M. Marklund and M. Servin, Phys. Rev. D \textbf{63}%
, 124003 (2001).

\bibitem{brodinmarklund}  G. Brodin and M. Marklund, Phys. Rev. Lett. 
\textbf{82}, 3012 (1999).

\bibitem{Papadouplous2001}  D. Papadopoulos, N. Stergioulas, L. Vlahos and
J. Kuijpers, A\&A \textbf{377}, 701 (2001).

\bibitem{bmd2}  G. Brodin, M. Marklund and P. K. S. Dunsby, Phys. Rev. D 
\textbf{62}, 104008 (2000).

\bibitem{Balakin2003}  A. B. Balakin, V. R. Kurbanova and W. Zimdahl, J.
Math. Phys., \textbf{44}, 5120 (2003)

\bibitem{Servin2000}  M. Servin, G. Brodin, M. Bradley and M. Marklund,
Phys. Rev E,\textbf{\ 62}, 8493 (2000).

\bibitem{ignatev}  Yu G. Ignat'ev, Phys. Lett. A \textbf{320}, 171 (1997).

\bibitem{kallberg2004}  A. Källberg, G. Brodin and M. Bradley, Phys. Rev. D 
\textbf{70}, 044014 (2004).

\bibitem{bmd1}  M. Marklund, G. Brodin and P. K. S. Dunsby, Astrophys. J. 
\textbf{536}, 875 (2000).

\bibitem{Moortgat2003}  J. Moortgat and J. Kuijpers, A\&A \textbf{402}, 905 (2003).

\bibitem{Mosquera2002}  H. J. M. Cuesta, Phys. Rev. D \textbf{65}, 64009 (2002).

\bibitem{Isliker} H. Isliker, I. Sandberg, L. Vlahos, Phys. Rev. D \textbf{74}, 104009 (2006).

\bibitem{Moortgat2006}  J. Moortgat and J. Kuijpers, MNRAS 368 
, 1110 (2006).

\bibitem{Papadoupolus2002}  D. Papadopoulos, Class Quantum Grav. \textbf{19}, 2939 (2002).

\bibitem{MDB2000}  M. Marklund, P.K.S. Dunsby, and G. Brodin, Phys. Rev. D 
\textbf{62}, 101501 (2000).

\bibitem{Hogan2002}  P. A. Hogan and E. M. O'Shea, Phys. Rev D \textbf{65},
124017 (2002).

\bibitem{kuiroukidis} A. Kuiroukidis, K. Kleidis, D. B. Papadopoulos and L. Vlahos, A\&A \textbf{471}, 409
(2007).

\bibitem{servinbrodin}  M. Servin and G. Brodin, Phys. Rev. D \textbf{68},
044017 (2003).

\bibitem{ALFA}  D. L. Jonews, R. J. Allen, J. P. Basart et al., Adv. Space
Res., \textbf{26, }743, (2000).

\bibitem{SIRA}  R. J. MacDowall, S. D.Bale, L. Dermaio, et al, Proceedings
of SPIE, \textbf{5659, }284 (2005).

\bibitem{Landau-Lifshitz}  L. D. Landau and E. M. Lifshitz, \textit{The Classical
Theory of Fields} (Pergamon Press, Oxford, 1975)




\bibitem{grishchuk}  L. P. Grishchuk and A. G. Polnarev, \textit{General
Relativity and Gravitation Vol. 2} ed. Held A (Plenum Press, New York, 1980)
pp 416-430.

\bibitem{doddeilbeck}  R. K. Dodd, J. C. Eilbeck, J. D. Gibbon and H. C.
Morris, \textit{Solitons and Nonlinear Wave Equations} (Academic Press,
London,1982).

\bibitem{Shukla} P. K. Shukla {\it Nonlinear waves, Chap. 11} Ed. L. 
Debnath.(Cambridge University Press, New York, 1983)













\end{thebibliography}
\end{document}